\documentclass[12pt]{article}
\usepackage[margin=1in]{geometry}

\usepackage{amsmath,amssymb,amsfonts}
\usepackage{setspace}

\usepackage[dvipsnames]{xcolor}
\usepackage{graphicx}
\usepackage{tikz}
\usepackage[nocompress]{cite}

\usepackage{authblk}

\usepackage{ulem} 
\newcommand{\ourG}{\mathbf{G}}
\newcommand{\ourB}{\mathbf{B}}

\usepackage{hyperref}
\hypersetup{colorlinks=true,allcolors=blue,urlcolor=cyan}

\begin{document}

\title{Natural modification of quantum uncertainty, modified gravity, and cosmology}

\author[1,2]{Christian G. B\"ohmer\footnote{Email: c.boehmer@ucl.ac.uk}}
\author[1,3]{Eissa Al-Nasrallah\footnote{Email: eissa.alnasrallah.24@ucl.ac.uk}}

\affil[1]{
  Department of Mathematics, University College London, \protect\\
  Gower Street, London WC1E 6BT, United Kingdom\protect\vspace{1ex}}
\affil[2]{
  Astrophysics Research Centre, School of Mathematics, Statistics \protect\\
  and Computer Science, University of KwaZulu-Natal, \protect\\
  Private Bag X54001, Durban 4000, South Africa\protect\vspace{1ex}}
\affil[3]{ Department of Physics, College of Science, Kuwait University, Sabah Al Salem University City, P.O. Box 5969, Safat 13060, Shadadiya, Kuwait}

\date{\today} 

\maketitle

\begin{abstract}
A common approach in physics and mathematics is to extend and modify theories and frameworks by considering what is often described as a ``natural'' extension or modification by including higher-order terms or by introducing other non-linearities. We show that such an approach must be taken with care as physical models can be connected in indirect ways. What looks like a natural approach in one setting will likely not be natural in another. We use the flat Friedmann-Lemaitre-Robertson-Walker equations of cosmology to connect the generalized uncertainty principle to modified theories of gravity. A simple additional term in one setting leads to enormous complications in the other. We identify Born-Infeld models as the only ones which appear natural in both settings.
\end{abstract}
\mbox{}\\[2ex]
\noindent ``This essay received an Honorable Mention in the 2026 Essay Competition of the Gravity Research Foundation.''
\thispagestyle{empty}
\clearpage

\doublespacing
\section{Introduction}
\setcounter{page}{1}
The entire structure of our physics knowledge is based on two firm foundations: general relativity (GR) and quantum field theory (QFT). Yet, the two theories are incompatible with each other, as general relativity is a purely classical theory. Moreover, we do not properly understand roughly 95\% of the energy content of the universe. This is made up of dark energy and dark matter. Attempts to reconcile the two theories have established the field of quantum gravity, which over the years, has introduced promising models like string theory and loop quantum gravity, yet none has presented a satisfactory theory of quantum gravity. While a modification of the current theories is inevitable, it remains an open question of how to correctly approach this.

Both GR and QFT have been well-tested within their relevant regimes, setting tight experimental constraints on different models. Thus, any modification in the theories has to necessarily reduce to the current forms in an appropriate limit. A safe approach often adopted is to introduce ``small'' modifications or perturbations to current theories such that they can be safely neglected within our tested frameworks but become significant in untested regimes. While one can find a plethora of research tackling the modifications and extensions of many theories, it is seldom investigated what such changes would imply in another setting. That is because each of GR and QFT have their own realms of application and thus small modifications remain irrelevant in the other theory. 

We challenge this philosophy in approaching modification, extensions, or generalizations of a theory that are often assumed to be ``small'' or ``natural''. Such a modification, which might well be small in one theory, can have drastic implications on the other, or at the least, appear quite ``unnatural''. We present a case for modifications often introduced in GR and QFT and show that they can be connected through cosmology and lead to unexpected results. Our approach relies on the cosmological implication of the modifications in each theory. 

Cosmological dynamics are described by the Friedmann equations, which for a flat Friedmann-Lemaitre-Robertson-Walker (FLRW) universe are given by
\begin{align}
    3H^2 = \kappa \rho \,, \qquad 2\dot{H} + 3H^2 = -\kappa p \,.
    \label{friedmann}
\end{align}
Here, $H=\dot{a}/a$ is the Hubble function, $a$ is the scale factor, $\kappa$ is the gravitational coupling constant, and $\rho$ and $p$ are the energy density and pressure, respectively. The dot denotes differentiation with respect to cosmological time $t$. A modification of the Einstein field equations yields a modified version of the Friedmann equation, which is often observed in modified gravity. The Friedmann equation can also be derived from a Hamiltonian, which is a function of the scale factor and its conjugate momentum. A changed canonical relation between these operators changes the Hamiltonian, and consequently the resulting Friedmann equation, by upgrading Heisenberg's uncertainty principle to the generalized uncertainty principle (GUP). With that, we will establish a relation between modified gravity and GUP using the cosmological Friedmann equations as a mediator between the two theories. 

\section{Modified gravity}

The gravitational forces experienced by matter are the effects of curvature in spacetime causing matter to move along geodesics accordingly. This curvature is determined by the Einstein field equations
\begin{align}
\label{eq_einstein_field}
    R_{\mu\nu}-\frac{1}{2}Rg_{\mu\nu} = \kappa T_{\mu\nu} \,,
\end{align}
where $R_{\mu\nu}$ is the Ricci tensor and $R$ is its trace, $g_{\mu\nu}$ is the metric, and $T_{\mu\nu}$ is the energy-momentum tensor. The field equations can be derived geometrically or through a least action principle using the Einstein-Hilbert action
\begin{align}
    \label{eq_EH_action}
    S_{\rm EH} = \frac{1}{2\kappa}\int \sqrt{-g}R \, d^4x + S_{\rm matter} \,.
\end{align}

Alternatives to GR have been proposed since the early days of GR through different techniques and approaches, which resulted in a large number of modified theories of gravity~\cite{Nojiri:2017ncd, Shankaranarayanan:2022wbx}. Geometrically, GR assumes the Levi-Civita connection which is metric compatible and symmetric. By lifting these constraints, one can obtain a theory that depends on torsion, the anti-symmetric part of the connection, or non-metricity, the covariant derivative of the metric. Even more generally, one can treat the metric and connection completely independently, resulting in metric-affine theories of gravity. Further approaches allow for higher-order theories, like $f(R)$ gravity, or non-minimal coupling to matter, scalar fields, or vector and tensor fields~\cite{Sotiriou:2008rp}. One could also increase the dimensionality of the spacetime or assume further symmetries like supersymmetry.

A simple approach to modifying gravity is to decompose the Ricci scalar $R$ into bulk $\ourG$ and boundary $\ourB$ terms~\cite{Boehmer:2021aji}. One can then write the Einstein action as
\begin{align}
\label{eq_einstein_action}
    S = \frac{1}{2\kappa}\int \sqrt{-g} (\ourG+\ourB) d^4x + S_{\rm m} \,,
\end{align}
where $\ourG$ and $\ourB$ are pseudo-scalars:
\begin{align}
\label{eq_B_G}
    \ourG = g^{\mu\nu}(\Gamma^\sigma_{\mu\lambda} \Gamma^\lambda_{\sigma\nu}-\Gamma^\sigma_{\mu\nu} \Gamma^\lambda_{\lambda\sigma})\,, 
    \qquad 
    \ourB = \partial_\sigma \Bigl[\sqrt{-g}\bigl( g^{\mu\nu}\Gamma^\sigma_{\mu\nu}-g^{\mu\sigma} \Gamma^\lambda_{\mu\lambda}\bigr)\Bigr]/\sqrt{-g} \,.    
\end{align}
Only the bulk term contributes to the derivation of the field equations~\eqref{eq_einstein_field} since the $\sqrt{-g}\ourB$ is a total derivative. The decomposition of the Ricci scalar in $f(R)=f(\ourG+\ourB)$ theories can be generalized to $f(\ourG,\ourB)$. Unlike $f(R)$ gravity, this theory is no longer diffeomorphism invariant and depends on the coordinates chosen. If one requires second-order equations, one must choose $f(\ourG)$, see~\cite{Boehmer:2021aji, Boehmer:2023fyl}.

In FLRW cosmology one obtains the cosmological equations of $f(\ourG)$ gravity to be
\begin{align}
    \label{eq_fG_friedmann}
    f(\ourG)+12H^2f'(\ourG) = 2\kappa \rho \,,
\end{align}
where one has $\ourG = -6H^2$. One of the strengths of $f(\ourG)$ gravity is that it links to $f(Q)$ gravity and $f(T)$ gravity. In the cosmological context, the result~\eqref{eq_fG_friedmann} is identical across these three approaches in the cosmological setting~\cite{Boehmer:2021aji, Boehmer:2023fyl}.

\section{Generalized uncertainty principle}

The Heisenberg uncertainty principle in quantum mechanics does not set a limit to the precision of determining either position or momentum separately. In other words, one can determine position with arbitrary large precision as long as uncertainty in momentum is allowed to be arbitrarily large. It is, however, a prediction of quantum gravitational physics that there exists a minimum measurable length of space~\cite{Hossenfelder:2012jw}. This can emerge for example from the Schwarzschild radius in black hole physics~\cite{Scardigli:1999jh} or the characteristic string length in String theory~\cite{Yoneya:2000bt}. The existence of a minimum measurable length would restrict the uncertainty of position measurements which requires a modification of Heisenberg's uncertainty principle to the so-called generalized uncertainty principle (GUP). In its simplest form~\cite{Kempf:1994su}, it is defined as
\begin{align}
    \label{eq_gup_uncertainty_relation}
    \Delta X \Delta P \geq \frac{\hbar}{2}\bigl(1+\beta (\Delta P)^2 \bigr)\,.
\end{align}
where $\beta$ is a positive constant, called the GUP parameter. This relation yields a modification of the canonical commutation relations:
\begin{align}
    \label{eq_commutation_relation}
    [X,P] = i\hbar \bigl(1+\beta P^2 \bigr)\,.
\end{align}
The modified commutation relation allows us to define a corresponding position and momentum operators in terms of Heisenberg's canonical operators as
\begin{align}
    \label{eq_gup_operators}
    X = x \,, \qquad P = p \bigl(1+\lambda_1 p +\lambda_2 p^2+\mathcal{O}(p^3) \bigr)\,.
\end{align}

In cosmological applications, one commonly considers the position operator as the scale factor $a$ and its corresponding conjugate momentum $p_a$. For the flat FLRW cosmology, the Hamiltonian is 
\begin{align}
    \label{eq_flrw_hamiltoian}
    \mathcal{H}_{\rm FLRW}(a,p_a) = N\frac{\kappa}{12}\frac{p_a^2}{a}- N\rho a^3+\epsilon\Pi \,.
\end{align}
Here $N$ is the lapse function typically set to $N=1$, $\Pi$ is its conjugate momentum, $\rho$ is the density of the universe, and $\epsilon$ is a coupling coefficient. The resulting Friedmann equation from~\eqref{eq_flrw_hamiltoian} is $H^2 =\kappa\rho/3$, identical to~\eqref{friedmann}.

Next, considering a GUP model with modified operators~\eqref{eq_gup_operators},  one would necessarily obtain a changed Friedmann equation~\cite{Aghababaei:2021gxe}. Given that we can also derive the Friedmann equations from the gravitational field equations from modified gravity, by choosing $f$ in~\eqref{eq_fG_friedmann}, we now establish the connection between the two theories.   

\section{Connecting modified gravity and GUP}

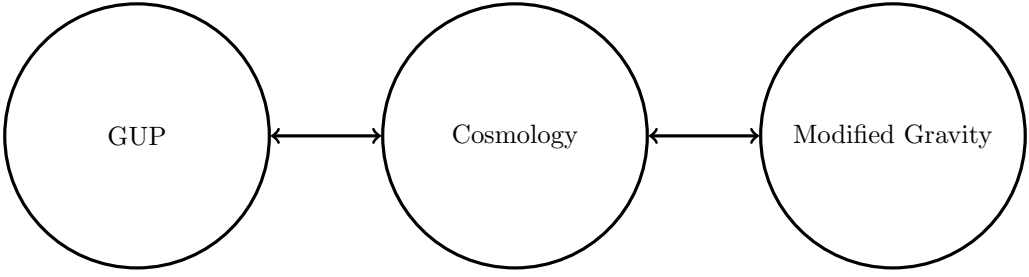
\begin{figure}[!htb]
    \centering
        \begin{tikzpicture}[
             node distance=5cm,
             every node/.style={circle, draw, minimum size=3.5cm}]
                \node[circle, draw, very thick] (cosmology)  {Cosmology};
                \node[circle, draw, very thick, left of=cosmology] (gup)  {GUP};
                \node[circle, draw, very thick, right of=cosmology] (modgr)  {Modified Gravity};
                \draw[<->][very thick] (gup)--(cosmology);
                \draw[<->][very thick] (cosmology)--(modgr);
        \end{tikzpicture}
    \caption{Connecting the theories together.}
    \label{fig:scheme}
\end{figure}

The Friedmann equations can be derived using two approaches, modified gravity and GUP. In both approaches, the foundational models of GR and Heisenberg's uncertainty principle are generalized to broader models. Instead of following the common approach of finding a natural generalization that is simple and remains as close to the original model as possible, we look for modified models that facilitate the transition between different theories as depicted in Figure~\ref{fig:scheme}. We thus proceed to develop the mechanism that connects the different theories.

First, we start with~\eqref{eq_fG_friedmann} and re-write as
\begin{align}
\label{eq_Friedmann_fG}
    \kappa\rho=-\ourG^{3/2}\frac{d}{d\ourG}\bigl[f(\ourG)\,\ourG^{-1/2} \bigr] :=\mathcal{K}(\ourG)\,,
\end{align}
where we define the right-hand side to be the function $\mathcal{K}(\ourG)$. One can think of this as a modified Hubble function. For a general Friedmann equation with given $\mathcal{K}(\ourG)$, one can formally solve for $f(\ourG)$ using integration
\begin{align}
    \label{eq_fG_eqn}
    f(\ourG) = -\ourG^{1/2}\int^\ourG \mathcal{K}(y)y^{-3/2}dy \,.
\end{align}
Thus, any given $f(\ourG)$ model leads to a Friedmann equation by using~\eqref{eq_fG_friedmann} and by substituting $\ourG=-6H^2$. Conversely, any given Friedmann equation can lead to a concrete $f(\ourG)$ modified gravity model by making that same substitution and solving~\eqref{eq_fG_eqn}.

A similar procedure can also be developed between the GUP-modified Hamiltonian and the Friedmann equations. The Hamiltonian in~\eqref{eq_flrw_hamiltoian} can be generalized to
\begin{align}
    \label{eq_hamiltonian_u}
    \mathcal{H}(u) = a^3\mathcal{F}(u) - \rho a^3+\epsilon\Pi \,,    
\end{align}
where we defined the variable $u=p/a^2$. In the case of Eq.~\eqref{eq_flrw_hamiltoian}, we simply have $\mathcal{F}(u)=\kappa u^2/12$. The choice of $u$ here is important to maintain a suitable relation between $p_a$ and $a$ that leads to the desired form of the Friedmann equation that only depends on the Hubble function. Since we are interested in models with Friedmann equations that can be connected to $f(\ourG)$ gravity, we find that this condition restricts our choice of functions $\mathcal{F}(p_a,a)$ to $\mathcal{F}(p_a/a^2)=\mathcal{F}(u)$.

Starting with Hamilton's equations, the Hubble parameter is derived as $H=\dot{a}/a=\partial \mathcal{F}(u)/\partial u$, while the equation for the conjugate momentum of the lapse function $\dot{\Pi}$ gives $\rho=\mathcal{F}(u)$ or equivalently $u=\mathcal{F}^{-1}(\rho)$. Those two relations yield the Friedmann equation using the inverse function theorem
\begin{align}
\label{eq_Hubble_GUP}
    H=\frac{\partial \mathcal{F}}{\partial u} = 
    \frac{\partial \mathcal{F}}{\partial (\mathcal{F}^{-1}(u))} = 
    \frac{1}{\bigl(\partial u/\partial \rho \bigr)} \,.
\end{align}
Thus, given a Hamiltonian~\eqref{eq_hamiltonian_u}, by identifying $\mathcal{F}(u)$ as the density $\rho$ and solving for $u(\rho)$, Eq.~\eqref{eq_Hubble_GUP} gives the relevant Friedmann equation. The reverse would be to use $H(\rho)$ from the Friedmann equation in~\eqref{eq_Hubble_GUP}, and integrate
\begin{align}
\label{eq_u_integral}
    u(\rho)=\int^\rho \frac{dy}{H(y)} \,,
\end{align}
to get $\mathcal{F}(u)$ for the modified Hamiltonian expression~\eqref{eq_hamiltonian_u} from the inverse of~\eqref{eq_u_integral}, where $\rho(u)=\mathcal{F}(u)$. 

If we consider the case for a flat FLRW universe with the Heisenberg uncertainty principle and GR, we have $\mathcal{F}(u)= \kappa u^2/12$. Setting $\rho=\mathcal{F}(u)$ and solving for $u$ gives $u=\sqrt{12\rho/\kappa}$. By taking the derivative with respect to $\rho$ and substituting in~\eqref{eq_Hubble_GUP} we get the well-known expression for Friedmann equation $H^2=\kappa\rho/3$. Then, substituting $H^2=-\ourG/6$ and evaluating~\eqref{eq_fG_eqn} gives $f(\ourG)=\ourG$ as expected for GR.

We see that the Friedmann equation for a flat FLRW universe has become an intermediary between two theories, namely GUP depicted by modified Hamiltonian and modified gravity depicted by $f(\ourG)$ models. A modified model given in any of these theories can be transferred to the other two using the above description, see Figure~\ref{fig:scheme}. This opens the door for a novel comparison between the modifications of the theories in which natural modifications of one theory are not guaranteed to lead to natural or expected modifications in the other two. Mathematically, this is because the derivative of an inverse function is difficult to relate back to the original function. 

The Hamiltonian deformation should be interpreted in an effective mini-superspace sense. Standard GUP models begin from a modified quantum commutator, whereas our construction asks which effective Hamiltonian deformation reproduces a given modified Friedmann equation. For a generic function $\mathcal{F}(u)$, this does not uniquely determine an underlying closed quantum algebra or deformed Poisson bracket. Thus, the connection to GUP should be understood phenomenologically.

\section{Natural and unnatural models}

Our formalism can now be utilized as a tool to investigate a variety of models. First, it can quickly be noticed that many modified Hamiltonian models that are considered simple, yield complicated expressions in the Friedmann equation and $f(\ourG)$ gravity. One can see that models of the form $\mathcal{F}(u)=u^2+u^3$ or $\mathcal{F}(u)=u^2+u^4$ immediately produce complicated expressions in the Friedmann equation and $f(\ourG)$ gravity due to the root structure. Likewise, the same can be seen by starting with $f(\ourG)=\ourG+\ourG^2$. This should not come as a surprise since the change of theory settings requires the integration of the reciprocal of an inverse function, see~\eqref{eq_fG_eqn} and~\eqref{eq_u_integral}, which simply complicates the resulting expressions. Thus, it is worth revisiting those models which are often thought to present the most natural extension of the current theories and investigate how the introduced changes modify the other theory.

An interesting model in this context is motivated by loop quantum cosmology~\cite{Ashtekar:2008zu} where the Friedmann equation is given by $3H^2=\kappa\rho (1-\rho/\rho_c )$. Using~\eqref{eq_Hubble_GUP}, we solve for $u$:
\begin{align}
    u=\sqrt{\frac{3}{\kappa}}\int\limits_{0}^\rho\frac{dy}{\sqrt{y(1-y/\rho_c)}}=2\sqrt {\frac{3}{\kappa}}\sqrt{\rho_c}\arcsin{\bigl(\rho/\rho_c\bigr)}\,.
\end{align}
Now solving for $\rho=\mathcal{F}(u)$ gives
\begin{align}
    \mathcal{F}(u)= \rho_c\sin^2\Bigl(\frac{1}{2}\sqrt{\frac{\kappa}{3\rho_c}}u \Bigr)\,.
\end{align}
This result is simple and probably expected in view of the form of the holonomies~\cite{Ashtekar:2005qt}. However, it would have been unlikely to make this choice independently of loop quantum gravity. Notice that an expansion around $u=0$ would return $\mathcal{F}(u)=\kappa u^2/12+\mathcal{O}(u^4)$ term, as one would expect. The situation changes quite dramatically in modified gravity. Since the Friedmann equation involves linear and quadratic terms in the density, one begins by substituting $\ourG$ for the Hubble parameter and solving the quadratic equation to get an expression $\kappa\rho=\mathcal{K}(\ourG)=\kappa\rho_c(1-\sqrt{1+2\ourG/\kappa\rho_c})/2$. Performing the integration in~\eqref{eq_fG_eqn} gives
\begin{align}
\label{eq_fG_loop}
    f(\ourG) = \kappa\rho_c\Biggl(1-\sqrt{1+\frac{2\ourG}{\kappa\rho_c}}+\sqrt{\frac{2\ourG}{\kappa\rho_c}}\sinh^{-1}\Bigl(\sqrt{\frac{2\ourG}{\kappa\rho_c}} \Bigr) \Biggr)\,.
\end{align}
Now this term is far from intuitive and more complicated than any modification one would be likely to introduce to an $f(\ourG)$ gravity model. While the term does reduce to $\ourG$ in the limit $\rho_c \to \infty$, it is clearly not a natural extension of general relativity. Thus, we obtain an expression for $f(\ourG)$ gravity and a modified Hamiltonian that are arguably unnatural, in spite of the expression leading to them being a well motivated quantum cosmology model. 

An interesting case in which a simple modification leads to a simple modification in the respective theory is the addition of a constant or a linear term. For a general expression of the quadratic Hamiltonian
\begin{align}
\label{eq_quadratic_Fu}
    \mathcal{F}(u) = \frac{\kappa}{12}u^2\bigl(1+\frac{2\gamma}{u}+\frac{4\gamma^2-\alpha^2}{4u^2} \bigr) \,,
\end{align}
where $\gamma$ is a constant. This yields a Friedmann equation and $f(\ourG)$ model with an additional constant term
\begin{align}
    H^2 = \frac{\kappa}{3}\rho + \frac{\alpha^2\kappa^2}{144}\,, 
    \qquad 
    f(\ourG) = \ourG - \frac{\alpha^2\kappa^2}{24} \,.
\end{align}
We notice that the parameter $\gamma$ is arbitrary in the $\mathcal{F}(u)$ expression and does not contribute to the modification of Friedmann equation or $f(\ourG)$ equation. The constant term in the Friedmann equation is particularly interesting as it is the cosmological constant in the $\Lambda$CDM model. This is expected for $\gamma=0$ and gives the correct cosmological constant, but it is surprising that the argument holds for $\gamma\neq0$. A comparison with the measured accelerated expansion rate of the universe can be used to bound the value of $\alpha$ which can further restrict models with modified Hamiltonian or $f(\ourG)$ gravity. Using $\Lambda \approx 10^{-122}\ell_p^{-2}$ one finds immediately $\alpha \approx 10^{-61} \ell_p^{-1}$.

The first square root term in~\eqref{eq_fG_loop} is similar to terms appearing in Born-Infeld type approaches~\cite{Born:1934gh, Born:1934dia}. Born-Infeld models are of particular interest as they can resolve singularity problems. We set
\begin{align}
    f(\ourG) = \lambda\Bigl(\sqrt{1+\frac{2\ourG}{\lambda}}-1 \Bigr)\,,
\end{align} 
where $1/\lambda \rightarrow 0$ leads to GR. It is extremely exciting that the corresponding function $\mathcal{F}(u)$ also produces a Born-Infeld like term, namely
\begin{align}
    \mathcal{F}(u) = \frac{\lambda}{2\kappa}\Bigl(\sqrt{1+\frac{\kappa^2}{3\lambda}u^2}-1 \Bigr) \,.
\end{align}
Terms of this form appear to be the only ones which maintain their mathematical structure in either setting. Hence, we find that Born-Infeld type theories are the only natural models to retain their form using this mechanism.

Our mapping depends on the matter model used to close the cosmological equations. We work with a single perfect fluid, so that the background dynamics can be written in terms of a single relation $H=H(\rho)$, allowing the construction of the key integral~(15). For multi-fluid cosmologies or scalar fields, this reduction may change its form, become non-invertible, or lose the structural simplicity found here. Consequently, the Born–Infeld result should be read as a naturalness statement within the restricted flat FLRW model, single-fluid setting considered in this essay, not as a universal matter-independent property.

The approach presented in this essay has connected models from separate theories that are often studied in isolation. This connection sheds light on an approach that is often implicitly used, that of simple or natural extensions of models. Simplicity has long been a guiding principle in physics. Here, we do not question the concept within the context of a certain theory, where it is often accepted naturally, but rather what it means across distinct theoretical frameworks. A natural and simple modification of a model can be well motivated and reasonable within the framework of a theory. We demonstrate that a connection can be established using the flat FLRW universe in which this same modification can imply unexpected and unnatural modifications in other settings. This challenges the very meaning of simplicity as a guiding principle, and we propose to focus on models that preserve their structure when changing frameworks.

What has been developed here should be understood as a background-level statement within the flat FLRW model. Matching the modified Friedmann equation does not establish a full equivalence between a phase-space deformation and a modified theory of gravity. In particular, we expect our results to change once inhomogeneous perturbations, additional gravitational degrees of freedom, tensor propagation, or ultraviolet matter dynamics are included.

In cases when a connection can be identified between models in the two theories, one can search for a deeper connection between them like the duality observed in~\cite{Nenmeli:2021orl}. Then one can extend the work presented here to cases in which the cosmology is less idealized than FLRW metric.
We hope that the work presented here will encourage colleagues to question the concepts of naturalness and simplicity as they embark on them in their work and consider what these actually mean in a broader scientific context. Namely, what motivates seeking simplicity if it is model dependent? Whether one chooses to abandon such guiding principles or to seek out alternative approaches is left to the reader. 

\subsection*{Acknowledgments}

Eissa Al-Nasrallah acknowledges funding provided by Kuwait University through its graduate student scholarship program.

\end{document}